\documentclass[runningheads,a4paper]{llncs}

\usepackage{amssymb}
\usepackage{graphicx}

\usepackage{url}
\newcommand{\keywords}[1]{\par\addvspace\baselineskip
\noindent\keywordname\enspace\ignorespaces#1}

\usepackage[
		ngerman	% TODO umstellen auf englisch oder unnötig?
		,colorinlistoftodos
		]{todonotes} %\todo, \missingfigure und \listoftodos

\usepackage{algorithmic}
\usepackage{algorithm}
\floatname{algorithm}{Algorithm}

\usepackage{lmodern}% http://ctan.org/pkg/lmodern
\usepackage{color}
\usepackage{multirow}
\usepackage[hidelinks]{hyperref}

\usepackage{booktabs}
\usepackage{units}

\usepackage{caption}
\usepackage{subcaption}

\usetikzlibrary{shapes,snakes}

\newcommand{\refalg}[1]{Algorithm~\ref{#1}}
\newcommand{\barabasi}{Barab{\'a}si-Albert}

\definecolor{blue1}{RGB}{31,119,180}
\definecolor{orange1}{RGB}{255,127,74}
\definecolor{green1}{RGB}{44,160,44}
\begin{document}

\mainmatter  % start of an individual contribution

\title{Dynamic Clustering in Social Networks using Louvain and Infomap Method}

\titlerunning{Dynamic Clustering using Louvain and Infomap}

\author{Pascal Held\and Benjamin Krause\and Rudolf Kruse}
\authorrunning{Held, Krause, Kruse}

\institute{Otto von Guericke University of Magdeburg\\
Department of Knowledge Processing and Language Engineering\\
Universit\"atsplatz 2, 39106 Magdeburg, GERMANY\\
\url{pascal.held@ovgu.de}\\
\url{benjamin.krause@st.ovgu.de}\\
\url{rudolf.kruse@ovgu.de}\\
\url{http://fuzzy.cs.uni-magdeburg.de}}


\toctitle{Lecture Notes in Computer Science}
\tocauthor{Authors' Instructions}
\maketitle

\begin{abstract}
%The abstract should summarize the contents of the paper and should contain at least 70 and at most 150 words. It should be written using the \emph{abstract} environment.

Finding communities or clusters in social networks is a famous topic in social network analysis.
Most algorithms are limited to static snapshots, so they cannot handle dynamics within the underlying graph.

In this paper we present a modification of the Louvain community detection method to handle changes in the graph without rerunning the full algorithm.
Also, we adapted the Louvain greedy approach to optimize the Infomap measure.

The main idea is, to recalculate only a small area around the changes. 
Depending on the graph size and the amount of changes, this yields a massive runtime decrease.

As validation data, we provide a graph generator, which produces specific community structures, at given times and also intermediate steps to transform the graph from one to another specific graph.

Experiments show that runtime decrease is possible without much loss of quality. These values depend on the reprocessed area inside the graph.

\end{abstract}

\section{Introduction}
\label{sec:introduction}

In the last years social networks became very popular. 
This is not limited to large public social networks like Facebook or Twitter.
There are also a bunch of hidden networks with social network characteristics like communication networks~\cite{Klimt2004}, co-authorship networks~\cite{Leskovec2006} or structure networks e.g.\ in websites~\cite{Leskovec2008}.
Also in human beings are such networks, like protein-protein-interaction networks~\cite{Han2004} or relations between different brain regions~\cite{Rubinov2010}.

Nowadays, social network analysis is a broad research area with different topics.
Some of them are analysis of social relations, special characteristics of single entities in groups, or density based analysis of different sub graphs.
Another core area is the analysis of community structure within the network~\cite{fortunato2010community}, which is focused in this paper.

Most approaches are based on the assumption that the network structure is static. Alternatively, different time stamps can be compared. 

If we look on interaction networks~\cite{Golder2007}, like \emph{comment}, \emph{share}, or \emph{like} activities on Facebook, we will find continuously changing networks.

In this paper we will present an algorithm, which can handle changes in the network. 
To do so, we change only some parts of the existing partitioning without running the whole algorithm again. 
This yields improved runtime.

\section{Terminology}
\label{sec:Terminology}

First, we give a brief introduction into graphs and community detection in social networks.

\subsection{Graph}
\label{ssec:graph}

Networks of information of any kind can be modeled as graphs.
A graph is an ordered tuple $G = (V, E)$, where $V$ is a set of unique vertices or nodes and $E \subseteq V\times V$ is a set of edges or links~\cite{biggs1976graph}.

If $\forall_{v_1, v_2 \in V}: (v_1, v_2) \in E \Rightarrow (v_2, v_1)  \in E$ holds true, the graph is undirected, otherwise directed.

A \emph{weighted} graph changes the definition of $E$ to a set of tuples $((u, v), w)$, where $(u, v)$ is the already defined 2-element subset of $V$ and $w$ is the weight of the edge, specifying its \emph{importance}~\cite{fortunato2010community}.
These edges can be referred to via the weight matrix $A$, where $A_{i,j}$ is the weight of the edge between node $i$ and node $j$.

For dynamic graphs we assume having a list of static graphs for a given time interval~\cite{holme2012temporal}.
Each of these static graphs or \textit{graph steps} represent a time step in that interval.
If you take two consecutive steps and compare nodes and edges, only a few changes relative to the size of the graphs should be seen.

\subsection{Community Structure and Clustering}
\label{ssec:communityclustering}

To find clusters, groups of nodes that are \emph{associated} in some way, is one of the main tasks of graph analysis.
Given a graph $G = (V, E)$, there is a subgraph $C = (V_c, E_c)$ with $V_c \subseteq V$ and $E_c \subseteq E$.
The \emph{internal degree} of $C$ is the number of edges that lead from one node in $C$ to another node in $C$.
The external degree on the other hand is the number of edges from nodes in $C$ to  the rest of the graph.
These values can hint to $C$ being a \emph{good} cluster.
If the internal degree is very high and the external degree is very low, $C$ could be an appropriate cluster~\cite{fortunato2010community}.

The main task is, to find a clustering, meaning a division in clusters, so that every node is assigned to one cluster.
This clustering should consider the communities present in the graph as good as possible.
There are different methods for measuring the quality of a clustering.
In this work, we mainly focus on the modularity~\cite{blondel2008fast} and the infomap~\cite{rosvall2008maps} measures.

For dynamic graphs, the task is quite similar~\cite{aston2014community}.
Fore every step, a partition is needed.
There are several other subtasks, like finding communities which are given implicitly in several steps and might not be visible in a static snapshot.
Furthermore, tracing and recognizing communities in dynamic graphs is a important but difficult task and won't be considered in this work.

\section{Related Work}
\label{sec:relatedwork}

Two well-known community detection algorithms are the Louvain method which is based on modularity and the Infomap algorithm.

\subsection{Louvain Clustering}
\label{ssec:louvain}

The Louvain algorithm~\cite{blondel2008fast} is a greedy agglomerative hierarchical Clustering approach which utilizes the modularity measure.
It was originally designed for unweighted, undirected graphs but can easily be adapted to weighted and directed graphs.
\autoref{equ:modularity} shows the calculation of the modularity measure, where $n_c$ is the number of clusters, $l_c$ is the number of intra-cluster edges, $d_c$ is the sum of degrees of all nodes in $c$ and $m$ is the number of edges in the graph.

\begin{equation}
	Q = \sum \limits^{n_c}_{c=1} \left[\frac{l_c}{m} - \left(\frac{d_c}{2m}\right)^2\right]
\label{equ:modularity}
\end{equation}

In this method, the partition is initialized with every node in its own cluster.
Then, for each node the modularity gain for shifting it to neighboring clusters is computed.
The largest positive gain is chosen and the node is moved.
This is done until no node is moved in a full iteration.
Then, the graph is modified in a way that every cluster is merged into a single node, while intra-cluster edges are added as loops and inter-cluster edges between the same clusters are merged into a single edge and have there weights added.
This induced graph is used in the next iteration, until there is no gain to get by shifting another node.
The modularity ranges between -1 and 1.
A value of 1 would represent the perfect clustering with no edges between clusters and all clusters densely connected. Lower measure hints to a worse result, while values below 0 indicate a really terrible clustering and are rarely reached.

To compute the gain $\Delta Q$ of moving a node $i$ into the cluster $C$, not the new modularity of the whole clustering is computed, a local value which represents the gain can be computed more efficiently instead, as shown in \autoref{equ:modularitylocal}.

\begin{equation}
	\Delta Q = \left[ \frac{\sum_{in} + k_{i,in}}{2m} - \left( \frac{\sum_{tot} + k_{i}}{2m} \right)^2 \right] - \left[ \frac{\sum_{in}}{2m} - \left( \frac{\sum_{tot}}{2m} \right)^2 - \left( \frac{k_i}{2m} \right)^2 \right]
\label{equ:modularitylocal}
\end{equation}

Here, $\sum_{in}$ is the sum of the weights of the links inside the cluster $C$, $\sum_{tot}$ is the degree of the cluster, $k_i$ is the degree of the node $i$, $k_{i, in}$ is the sum of the weights of links from $i$ to nodes in $C$.

\subsection{Infomap Clustering}
\label{ssec:infomap}

The Infomap measure was designed to measure the quality of a clustering by estimating the length of a code for paths through the graph utilizing the probability flow of random walks~\cite{rosvall2008maps}.
In contrast to maximizing modularity, the fundamental approach of Infomap clustering is to utilize flows in the graph.
This works well if clusters have a connected flow inside, meaning that if you randomly follow the direction of edges, you tend to stay inside the same cluster. 
It does not work if there is little or no flow at all or if a random walk ends up in deadlocks a lot.

In this work, we used an approximation to this Infomap value as described in~\cite{van2013graph}.
\autoref{equ:wlogv} shows the formula used.
The volume $v_c = \sum_{i \in c}s_i$ of a cluster $c$ is the sum of the strengths $s_i = \sum_{j \in V} a_{ij}$, the sum of all edge weights $a_{ij}$ incident to that node, of all nodes in $c$. 
The within weight $w_c = \sum_{i,j \in c}a_{ij}$ is the sum of all intra-cluster edge weights.
The total volume $M = v_V$ of the graph is the sum of the volume of all nodes in the graph, which is twice the sum of all edge weights for weighted and twice the number of edges for unweighted graphs.
The normalized within weight $\widehat{w}_c = w_c/M$ and the normalized volume $\widehat{v}_c = v_c/M$ of cluster $c$ are used to compute the actual \emph{wlogv} value.

\begin{equation}
	Q_{wlogv}(C) = \sum_{c \in C} \widehat{w}_c \log (\widehat{v}_c)
\label{equ:wlogv}
\end{equation}

%\subsubsection{Infomap gain}
%\label{sssec:infogain}

Since \autoref{equ:wlogv} consists of a sum over values specific for each cluster, the calculation of the gain of moving a node is easy to implement.
The overall gain of the clustering can be computed by just calculating the new values for the cluster the node was moved out of and the cluster it was moved into.
This is very similar to the way Louvain computes the gain of moving a node.

\section{Algorithm}
\label{sec:algorithm}

To adapt static clustering algorithms to dynamic tasks, we concentrated on using as much information from already processed steps as possible and iterate as few nodes and edges as possible, based on intelligently deciding which parts of the graphs are worth to be processed.

The basic idea is, to use the partition computed in the last step and minimize the number of iterated nodes by using information about what areas in the current graph were actually changed in the first place.
We initialize the algorithm with the last partition and optimize the procedure of iterating over the nodes to compute the current partition.
In the first iteration of the dendrogram generation, we only iterate nodes which are close to an edge which was changed from the last to the current graph.
Being \emph{near} can be interpreted as directly being adjacent to such an edge or having a path to an adjacent node which is shorter or equal to a given threshold.
This distance can be used to tweak the accuracy and the runtime of the algorithm.

Only iterating potentially relevant nodes leads to decreased complexity and gives a good base for the basic algorithm to work on.

The algorithm is based on the assumption that the dynamic graph has a relatively low amount of changes compared to the number of edges.
If every node is \textit{near} a changed edge, this algorithm doesn't provide any advantages.

\subsection{Modification}
\label{sssec:modification}

The initial graph is processed in the same way as in the static algorithm.
For every following step, a modified version of the last partition is used.
It is modified in that way, that labels of nodes which are \emph{close} to changes in the graph are deleted or the nodes are moved to their own separated cluster.
This refers to the standard Louvain initialization, where every node starts in its own cluster.
Changes are directly loaded via a specific file type we saved our graphs into.
Whether a node is \emph{close} depends on the \emph{delete range} chosen.
This range determines the length of the shortest path a node must have to a node directly involved in a change to be \emph{close}.
A delete range of zero only includes nodes, which are adjacent to an edge changed since the last step.
Any higher value increases the area, which is reset.

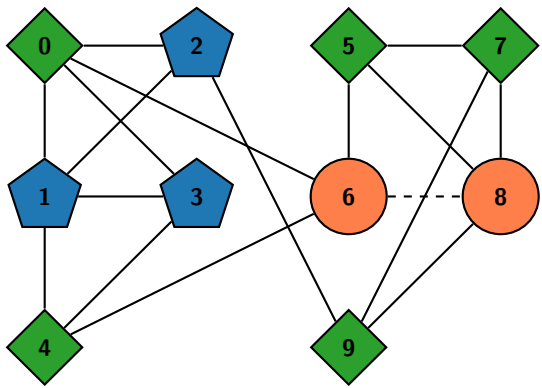
\begin{figure}[h]
		%\resizebox{\textwidth}{!}{	% ! -> proportional
		\centering
		\begin{tikzpicture}[node distance=2cm, thick,
		  some node/.style={circle,fill=gray!10,draw,font=\sffamily\small\bfseries},
		  diamond node/.style={diamond,fill=gray!10,draw,font=\sffamily\small\bfseries},
		  poly node/.style={regular polygon,regular polygon sides=5,fill=gray!10,draw,font=\sffamily\small\bfseries},
		  descr/.style={fill=white,inner sep=2.5pt}]	  
		  
		  \node[diamond node, minimum size=1cm, fill=green1] (0) {0};
		  \node[poly node, minimum size=1cm, fill=blue1] (1) [below of=0] {1};
		  \node[poly node, minimum size=1cm, fill=blue1] (2) [right of=0] {2};
		  \node[poly node, minimum size=1cm, fill=blue1] (3) [below of=2] {3};
		  \node[diamond node, minimum size=1cm, fill=green1] (4) [below of=1] {4};
		  
		  \node[diamond node, minimum size=1cm, fill=green1] (5) [right of=2] {5};
		  \node[some node, minimum size=1cm, fill=orange1] (6) [below of=5] {6};
		  \node[diamond node, minimum size=1cm, fill=green1] (7) [right of=5] {7};
		  \node[some node, minimum size=1cm, fill=orange1] (8) [below of=7] {8};
		  \node[diamond node, minimum size=1cm, fill=green1] (9) [below of=6] {9};
		  
		  \path[every node/.style={font=\sffamily\small}]
		    (0) edge  (1)
		        edge  (2)
		        edge  (3)
		    (1) edge  (2)
		        edge  (3)
		        edge  (4)
		    (3) edge  (4)
		    
		    (5) edge  (6)
		        edge  (7)
		        edge  (8)
		    (6) edge [dashed] (8)
		    (7) edge  (8)
		        edge  (9)
		    (8) edge  (9)
		    
		    (2) edge  (9)
		    (0) edge  (6)
		    (4) edge  (6);

		\end{tikzpicture}
		\caption{Example for the influence of delete range. A change happened between $6$ and $8$ (dashed  edge). Nodes to be iterated for delete range 0 are displayed as \textcolor{orange1}{orange circles}, nodes for delete range 1 are \textcolor{green1}{green diamonds} and nodes for delete range 2 are \textcolor{blue1}{blue pentagons}.}
		\label{fig:deleterange_example}
\end{figure}

\autoref{fig:deleterange_example} shows an example.
Given the edge $(6, 8)$ has changed since the last step.
Whether the edge was added, deleted or simply had it's weight changed does not matter.
For a delete range of 0, the set of nodes to be processed only contains $6$ and $8$, the nodes directly involved in the change.
If the delete range is increased to 1, the nodes $6$, $8$, and their neighbors are added, which leads to the set
$\lbrace 0, 4, 5, 6, 7, 8, 9\rbrace$.
A delete range of 2 furthermore adds the nodes $1$, $2$ and $3$.
This is only a small-scaled example to show what influence the delete range has on the area which is to be examined.

The modified partition is passed to the algorithm with a list of all affected nodes.
In the first iteration of the algorithm, only these nodes are processed.
This yields to a massive speed boost, because the first level of the dendrogram, where the most nodes have to be iterated, takes most of the processing time.
By reducing the number of these nodes, the time conserved can be substantial.
The following levels of the dendrogram consider the full graph.
A higher delete range means that more nodes have to be processed on the first level and therefore the algorithm needs more time.

%\begin{figure}[htbp]
%	\begin{minipage}{0.45\textwidth}
%		\resizebox{\textwidth}{!}{	% ! -> proportional
%		\centering
%  		\includegraphics[width=0.9\textwidth]{images/deleterange0-localoptimum.png}
%		}
%		\caption{Modularity for Louvain and Louvain-dyn with delete range 0. A local optimum can't be left with a range this small.}
%		\label{fig:deleterange0localoptimum}
%	\end{minipage}
%	% Auffüllen des Zwischenraums
%	\hfill
%	\begin{minipage}{0.45\textwidth}
%		\resizebox{\textwidth}{!}{	% ! -> proportional
%		\centering
%  		\includegraphics[width=0.9\textwidth]{images/deleterange2-localoptimum.png}
%		}
%		\caption{Modularity for Louvain and Louvain-dyn with delete range 2. The local optimum can be overcome.}
%		\label{fig:deleterange0localoptimum}
%	\end{minipage}
%
%\end{figure}
%\todo{nebeneinander? oder rechtes bild garnicht?}

%\subsection{Noise Detection}
%\subsection{Dealing with Noise}
%\label{sssec:noisedetection}

For each algorithm we use the same method to deal with noise.
In order to detect noise clusters, we iterate over all clusters and determine their sizes.
Clusters with less than a minimum amount of nodes will have all their nodes transferred to a designated noise cluster.
The default number of nodes a cluster needs to persist is two.
That way, only single node clusters will be removed.

\section{Dynamic Graph Generation}
\label{sec:dynamicgraphgen}

To evaluate the novel methods, we designed a graph generator to produce dynamic graphs with controllable ground truth.
The idea is to generate graphs with a given number of clusters and a given number of nodes in these clusters.
To realize this, a graph generator was created, which creates single \emph{graph steps}, which serve as controllable base graphs, and computes the intermediate graphs via differences between two graph steps.
The whole procedure can be seen in \refalg{alg:gen_dyn_graph} and \ref{alg:gen_graph}.
Each graph step is assembled by creating a single \barabasi~graph for each of the desired clusters and merging these graphs to a single graph (see~\cite{barabasi1999emergence} and~\cite{held2015merging}), where each of the \emph{cluster graphs} serves as a clearly separated cluster (see \autoref{fig:graphgeneration_merge01} and \ref{fig:graphgeneration_merge02}).

\begin{figure}[htbp]
	\begin{minipage}{0.45\textwidth}
		\resizebox{\textwidth}{!}{	% ! -> proportional
		\begin{tikzpicture}[auto,node distance=2cm, thick,
		  some node/.style={circle,fill=gray!10,draw,font=\sffamily\small\bfseries},
		  main node/.style={circle,fill=gray!90,draw,font=\sffamily\small\bfseries},
		  adjacency node/.style={circle,fill=gray!70,draw,font=\sffamily\small\bfseries},
		  descr/.style={fill=white,inner sep=2.5pt}]
		
		  \node[some node, minimum size=1cm] (0) {0};
		  \node[some node, minimum size=1cm] (1) [below of=0] {1};
		  \node[some node, minimum size=1cm] (2) [right of=0] {2};
		  \node[some node, minimum size=1cm] (3) [below of=2] {3};
		  \node[some node, minimum size=1cm] (4) [below of=1] {4};
		  
		  \node[some node, minimum size=1cm] (5) [right of=2] {5};
		  \node[some node, minimum size=1cm] (6) [below of=5] {6};
		  \node[some node, minimum size=1cm] (7) [right of=5] {7};
		  \node[some node, minimum size=1cm] (8) [below of=7] {8};
		  \node[some node, minimum size=1cm] (9) [below of=6] {9};
		  
		  \path[every node/.style={font=\sffamily\small}]
		    (0) edge  (1)
		        edge  (2)
		        edge  (3)
		    (1) edge  (2)
		        edge  (3)
		        edge  (4)
		    (3) edge  (4)
		    
		    (5) edge  (6)
		        edge  (7)
		        edge  (8)
		    (6) edge  (8)
		    (7) edge  (8)
		        edge  (9)
		    (8) edge  (9);
		    
		\end{tikzpicture}
		}
		\caption{Two separately generated \barabasi~graphs}
		\label{fig:graphgeneration_merge01}
	\end{minipage}
	% Auffüllen des Zwischenraums
	\hfill
	\begin{minipage}{0.45\textwidth}
		\resizebox{\textwidth}{!}{	% ! -> proportional
		\begin{tikzpicture}[node distance=2cm, thick,
		  some node/.style={circle,fill=gray!10,draw,font=\sffamily\small\bfseries},
		  main node/.style={circle,fill=gray!90,draw,font=\sffamily\small\bfseries},
		  adjacency node/.style={circle,fill=gray!70,draw,font=\sffamily\small\bfseries},
		  descr/.style={fill=white,inner sep=2.5pt}]
		
		  \node[some node, minimum size=1cm] (0) {0};
		  \node[some node, minimum size=1cm] (1) [below of=0] {1};
		  \node[some node, minimum size=1cm] (2) [right of=0] {2};
		  \node[some node, minimum size=1cm] (3) [below of=2] {3};
		  \node[some node, minimum size=1cm] (4) [below of=1] {4};
		  
		  \node[some node, minimum size=1cm] (5) [right of=2] {5};
		  \node[some node, minimum size=1cm] (6) [below of=5] {6};
		  \node[some node, minimum size=1cm] (7) [right of=5] {7};
		  \node[some node, minimum size=1cm] (8) [below of=7] {8};
		  \node[some node, minimum size=1cm] (9) [below of=6] {9};
		  
		  \path[every node/.style={font=\sffamily\small}]
		    (0) edge  (1)
		        edge  (2)
		        edge  (3)
		    (1) edge  (2)
		        edge  (3)
		        edge  (4)
		    (3) edge  (4)
		    
		    (5) edge  (6)
		        edge  (7)
		        edge  (8)
		    (6) edge  (8)
		    (7) edge  (8)
		        edge  (9)
		    (8) edge  (9)
		    
		    (2) edge [line width=2.5pt, draw=orange1] (9)
		    (0) edge [line width=2.5pt, draw=orange1] (6)
		    (4) edge [line width=2.5pt, draw=orange1] (6);

		\end{tikzpicture}
		}
		\caption{Resulting graph after merging both \barabasi~graphs}
		\label{fig:graphgeneration_merge02}
	\end{minipage}
\end{figure}

\begin{figure}[h]
	\centering
  	\includegraphics[width=0.9\textwidth]{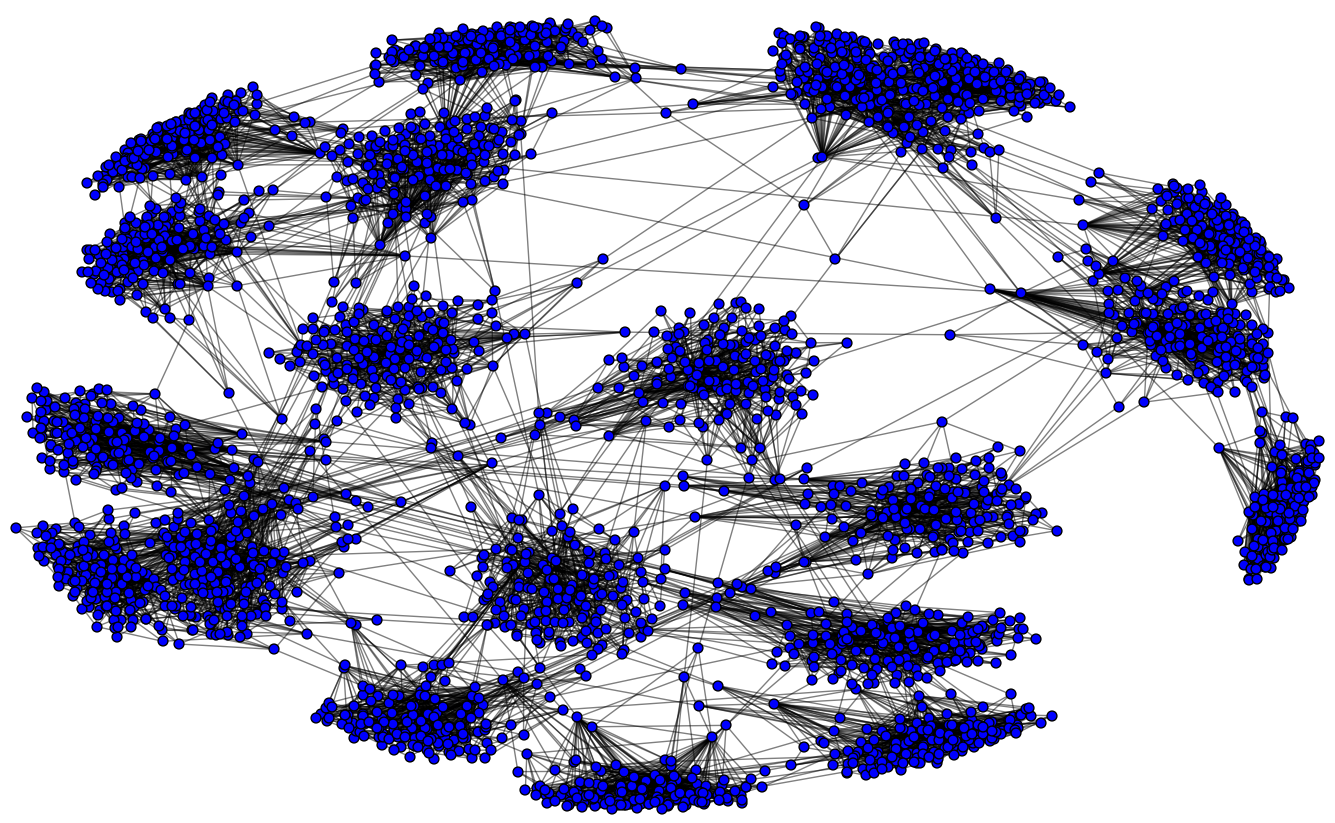}
	\caption{Example for a generated graph with 2000 nodes and 20 clusters}
	\label{fig:examplegraph01}
\end{figure}

The merge is executed with every two pairs of graphs, to receive graphs with two clusters in the first step.
These are recursively merged until a single graph with the requested number of clusters results.
\autoref{fig:examplegraph01} shows an example for a graph generated with this method.
%You can see how some pairwise clusters are more densely connected than others.
%This is the result of the pairwise merge, where some clusters are joining sooner than others and therefore receive more intercluster edges.

We used the minimal merge, which conserves the structure of both graphs merged and connects them with additional edges via preferential attachment.
This way, the desired structure, where each merged graph serves as a single, well-seperated cluster, can be achieved.
Due to preferential attachment the \barabasi~model is particularly suitable. It creates graphs with exponentially distributed node degrees, which serve great as a single cluster with hubs and border nodes.

Additionally, a number of intermediate steps between the predefined steps is given.
To fill these, the differences, namely edges to add or to delete, between two graph steps are computed.
The required changes are distributed between the intermediate time steps and applied one after another.
This way, each intermediate graph receives some of the changes needed to advance to the next time step.
When the last changes are applied, the resulting graph matches the target graph step.

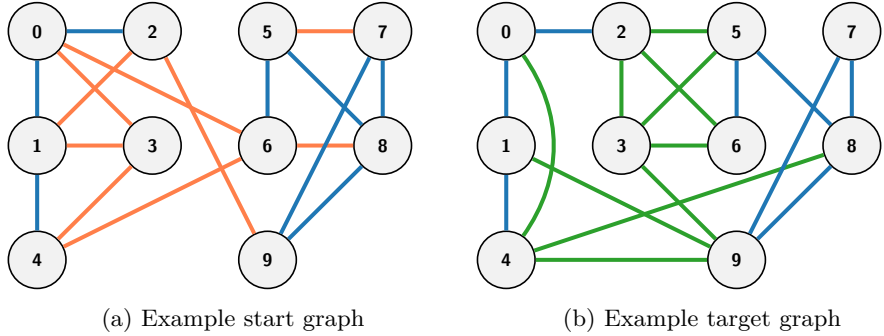
\begin{figure}
\begin{subfigure}[c]{0.5\textwidth}

\resizebox{0.9\textwidth}{!}{	% ! -> proportional
		\begin{tikzpicture}[auto,node distance=2cm, thick,
		  some node/.style={circle,fill=gray!10,draw,font=\sffamily\small\bfseries},
		  main node/.style={circle,fill=gray!90,draw,font=\sffamily\small\bfseries},
		  adjacency node/.style={circle,fill=gray!70,draw,font=\sffamily\small\bfseries},
		  descr/.style={fill=white,inner sep=2.5pt}]
		
		  \node[some node, minimum size=1cm] (0) {0};
		  \node[some node, minimum size=1cm] (1) [below of=0] {1};
		  \node[some node, minimum size=1cm] (2) [right of=0] {2};
		  \node[some node, minimum size=1cm] (3) [below of=2] {3};
		  \node[some node, minimum size=1cm] (4) [below of=1] {4};
		  
		  \node[some node, minimum size=1cm] (5) [right of=2] {5};
		  \node[some node, minimum size=1cm] (6) [below of=5] {6};
		  \node[some node, minimum size=1cm] (7) [right of=5] {7};
		  \node[some node, minimum size=1cm] (8) [below of=7] {8};
		  \node[some node, minimum size=1cm] (9) [below of=6] {9};
		  
		  \path[every node/.style={font=\sffamily\small}]
		    (0) edge [line width=2pt, draw=blue1] (1)
		        edge [line width=2pt, draw=blue1] (2)
		        edge [line width=2pt, draw=orange1] (3)
		    (1) edge [line width=2pt, draw=orange1] (2)
		        edge [line width=2pt, draw=orange1] (3)
		        edge [line width=2pt, draw=blue1] (4)
		    (3) edge [line width=2pt, draw=orange1] (4)
		    
		    (5) edge [line width=2pt, draw=blue1] (6)
		        edge [line width=2pt, draw=orange1] (7)
		        edge [line width=2pt, draw=blue1] (8)
		    (6) edge [line width=2pt, draw=orange1] (8)
		    (7) edge [line width=2pt, draw=blue1] (8)
		        edge [line width=2pt, draw=blue1] (9)
		    (8) edge [line width=2pt, draw=blue1] (9)
		    
		    (2) edge [line width=2pt, draw=orange1] (9)
		    (0) edge [line width=2pt, draw=orange1] (6)
		    (4) edge [line width=2pt, draw=orange1] (6);
		    
		\end{tikzpicture}
		}
	\subcaption{Example start graph}
	\label{fig:graphgeneration_phase01}

\end{subfigure}
\begin{subfigure}[c]{0.5\textwidth}
\resizebox{0.9\textwidth}{!}{	% ! -> proportional
		\begin{tikzpicture}[node distance=2cm, thick,
		  some node/.style={circle,fill=gray!10,draw,font=\sffamily\small\bfseries},
		  main node/.style={circle,fill=gray!90,draw,font=\sffamily\small\bfseries},
		  adjacency node/.style={circle,fill=gray!70,draw,font=\sffamily\small\bfseries},
		  descr/.style={fill=white,inner sep=2.5pt}]
		
		  \node[some node, minimum size=1cm] (0) {0};
		  \node[some node, minimum size=1cm] (1) [below of=0] {1};
		  \node[some node, minimum size=1cm] (2) [right of=0] {2};
		  \node[some node, minimum size=1cm] (3) [below of=2] {3};
		  \node[some node, minimum size=1cm] (4) [below of=1] {4};
		  
		  \node[some node, minimum size=1cm] (5) [right of=2] {5};
		  \node[some node, minimum size=1cm] (6) [below of=5] {6};
		  \node[some node, minimum size=1cm] (7) [right of=5] {7};
		  \node[some node, minimum size=1cm] (8) [below of=7] {8};
		  \node[some node, minimum size=1cm] (9) [below of=6] {9};
		  
		  \path[every node/.style={font=\sffamily\small}]
		    (0) edge [line width=2pt, draw=blue1](1)
		        edge [line width=2pt, draw=blue1] (2)
		        edge [line width=2pt, draw=green1, bend left=35] (4)
		    (1) edge [line width=2pt, draw=blue1] (4)
		        edge [line width=2pt, draw=green1] (9)
		    (2) edge [line width=2pt, draw=green1] (3)
		        edge [line width=2pt, draw=green1] (5)
		        edge [line width=2pt, draw=green1] (6)
		    (3) edge [line width=2pt, draw=green1] (5)
		        edge [line width=2pt, draw=green1] (6)
		        edge [line width=2pt, draw=green1] (9)
		    (4) edge [line width=2pt, draw=green1] (8)
		        edge [line width=2pt, draw=green1] (9)
		    (5) edge [line width=2pt, draw=blue1] (6)
		        edge [line width=2pt, draw=blue1] (8)
		    (7) edge [line width=2pt, draw=blue1] (8)
		        edge [line width=2pt, draw=blue1] (9)
		    (8) edge [line width=2pt, draw=blue1] (9)	    
		    ;

		\end{tikzpicture}
		}
		\subcaption{Example target graph}
		\label{fig:graphgeneration_phase02}
\end{subfigure}
\caption{Example for morphing from start to target graph. Constant edges are \textcolor{blue1}{blue} \{(0, 1), (0, 2), (1, 4), (5, 6), (5, 8), (7, 8), (7, 9), (8, 9)\}, deleted edges are \textcolor{orange1}{orange} \{(0, 3), (1, 2), (1, 3), (2, 9), (3, 4), (4, 6), (5, 7), (6, 8)\}, added edges are \textcolor{green1}{green} \{(0 ,4), (1, 9), (2, 3), (2, 5), (2, 6), (3, 5), (3, 6), (3, 9), (4, 8), (4, 9)\}.}
\label{fig:graphgeneration_phase}
\end{figure}

In  \autoref{fig:graphgeneration_phase} you can see an example for such a transition.
The clusters shift from two \emph{vertically} aligned clusters to one \emph{central} and one surrounding cluster.
Edges are colored referring to their state of change (constant, add, or delete).

This procedure is applied to all consecutive pairs of graph steps, until we get a dynamic graph, which has a graph step at the beginning and a number of sequences of a given number of intermediate graphs and a subsequent graph step.
The detailed algorithms for creating a dynamic graph and creating a single graph given a graph step can be seen in \refalg{alg:gen_dyn_graph} and \ref{alg:gen_graph}.

The resulting dynamic graph has some properties additional to the ones specified.
The first, last, and every predefined graph step are fairly easy to cluster, because the clusters are strongly separated.
The intermediate steps can be a little bit more challenging, because a good community structure is not guaranteed.
How hard the clustering exactly is is dependent on how different the two consecutive graph steps are and how many intermediate steps between predefined graphs there are.

\section{Experiments and Results}
\label{sec:experimentsresults}

Using the generated dynamic graphs as described in \autoref{sec:dynamicgraphgen}, we measured modularity, Infomap value, and time for each dynamic approach and its static base method.
The effects of the delete range parameter and different dynamic graph properties like number of nodes and number of changes per time step were analyzed.

\subsection{Experiment Setup}
\label{ssec:experiments}

Each graph step of the dynamic graph is clustered multiple times with each of the four algorithms (Louvain, Louvain-dyn, Infomap and Infomap-dyn).
For each algorithm, the average time is measured and evaluation values, Modularity and Infomap value, are computed from the resulting partitions.
This was done multiple times per graph to find an appropriate average.
%We will call this the \emph{base loop} from here on.

\autoref{tab:evaluationvalues} shows average evaluation values for a sample graph with 3000 nodes.
Depending on the delete range chosen, the time taken by the dynamic algorithms compared to the static ones ranges from about \unit[22]{\%} to about \unit[106]{\%}.
The evaluation values stay at about \unit[100]{\%} and for Infomap even tend to improve when compared to the static algorithm.
This shows how great the gain in complexity can be, without loosing much quality. 
Due to the greedy approach used by both base algorithms and the nature of the Infomap value, the usage of information collected in the previous step seems to improve the overall quality slightly.
The user has a wide array of options for choosing the parameter delete range.

%\begin{table}[]
%\centering
%\caption{Evaluation values for delete range 1 on a graph with 3000 nodes, 200 steps, 100 changes per step and 20 to 30 clusters with about equal size, respectively}
%\label{tab:evaluationvalues}
%\setlength{\tabcolsep}{2mm}
%\begin{tabular}{llll}
%\toprule
%~ & Time (in s) & Modularity & Infomap wlogv \\
%\midrule
%Louvain 				& 0.700 & 0.901 & 1.308 \\
%Louvain dynamic   & 0.584 (\unit[83.43]{\%}) & 0.896 (\unit[99.45]{\%}) & 1.235 (\unit[94.42]{\%}) \\
%\midrule
%Infomap					& 2.579 & 0.871 & 1.456 \\
%Infomap dynamic  & 1.388 (\unit[53.82]{\%}) & 0.878 (\unit[97.99]{\%}) & 1.462 (\unit[100.41]{\%}) \\
%\bottomrule 
%\end{tabular}
%\end{table}

\begin{table}[]
\centering
\caption{Percentage of dynamic evaluation values compared to static for delete ranges 0 to 3 on a graph with 3000 nodes, 2200 steps, 10 changes per step and 20 to 30 clusters with about equal size, respectively}
\label{tab:evaluationvalues}
\setlength{\tabcolsep}{2mm}
%\begin{tabular}{llll}
%\toprule
%~ & Time & Modularity & Infomap wlogv \\
%\midrule
%\textbf{Louvain} & ~ & ~ & ~ \\
%delete range 0 & 36,23758534 & 99,28323595 & 93,25142413 \\
%delete range 1 & 40,65227614 & 99,28218345 & 93,22443412 \\
%delete range 2 & 64,60811812 & 99,79663443 & 96,44132591 \\
%delete range 3 & 106,0178569 & 99,8561339 & 99,20436119 \\
%\midrule
%\textbf{Infomap} & ~ & ~ & ~ \\
%delete range 0 & 22,78145497 & 101,54135 & 96,22058106 \\
%delete range 1 & 30,64778825 & 101,2184379 & 100,3850431 \\
%delete range 2 & 57,68515178 & 100,4890487 & 100,4107361 \\
%delete range 3 & 86,99305754 & 99,99535174 & 99,85309678 \\
%\bottomrule 
%\end{tabular}
\begin{tabular}{lrrr}
\toprule
~ & Time & Modularity & Infomap wlogv \\
\midrule
\textbf{Louvain} & ~ & ~ & ~ \\
delete range 0 & \unit[36,24]{\%} & \unit[99,28]{\%} & \unit[93,25]{\%} \\
delete range 1 & \unit[40,65]{\%} &  \unit[99,28]{\%} & \unit[93,22]{\%} \\
delete range 2 & \unit[64,61]{\%} &  \unit[99,80]{\%} & \unit[96,44]{\%} \\
delete range 3 & \unit[106,02]{\%} & \unit[99,86]{\%} & \unit[99,20]{\%} \\
\midrule
\textbf{Infomap} & ~ & ~ & ~ \\
delete range 0 & \unit[22,78]{\%} & \unit[101,54]{\%} & \unit[96,22]{\%} \\
delete range 1 & \unit[30,65]{\%} & \unit[101,22]{\%} & \unit[100,39]{\%} \\
delete range 2 & \unit[57,69]{\%} & \unit[100,49]{\%} & \unit[100,41]{\%} \\
delete range 3 & \unit[86,99]{\%} & \unit[100,00]{\%} & \unit[99,85]{\%} \\
\bottomrule 
\end{tabular}
\end{table}

In the following we will describe experiments which analyze the influence of different parameters and graph properties.
In each case, the corresponding parameter was modified.

\subsection{Delete Range}
\label{ssec:experiments_deleterange}
\begin{figure}
	\centering
  	\includegraphics[width=0.8\textwidth]{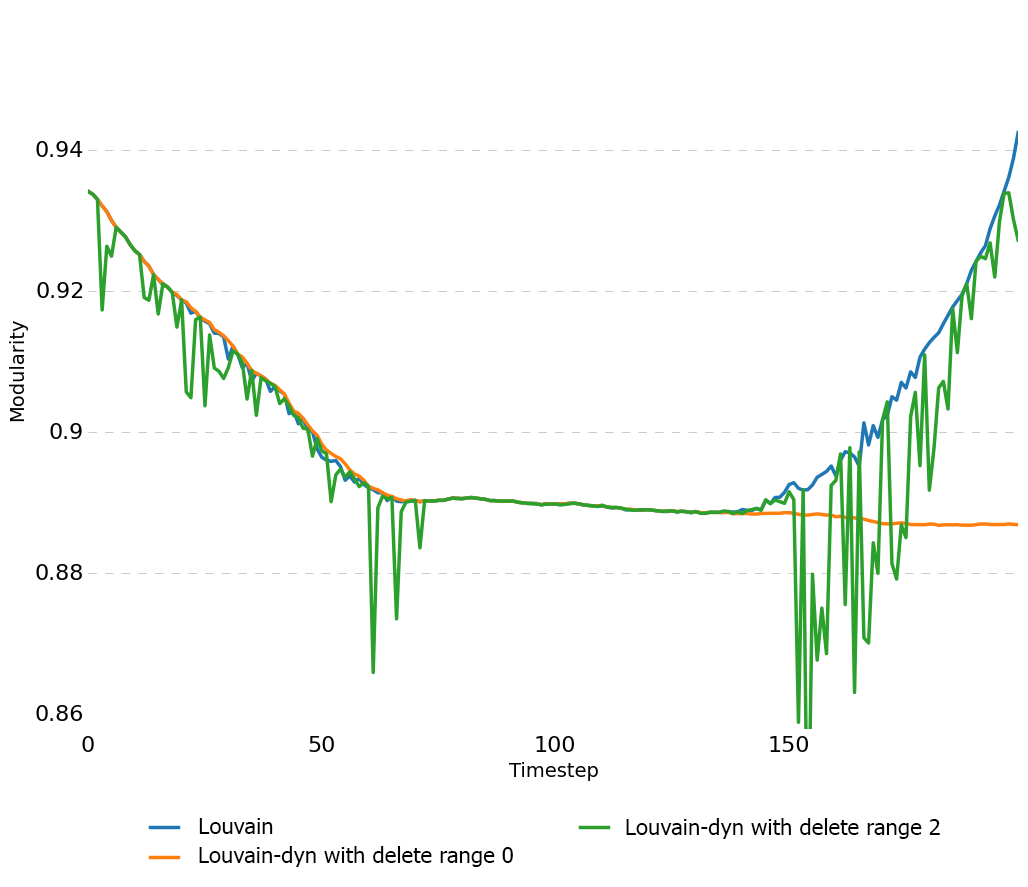}
	\caption{Modularity for Louvain and Louvain-dyn with delete range 0 and 2. The Louvain values for delete range 2 are identical. A local optimum can't be left with delete range 0, but it works with delete range 2.}
	\label{fig:deleterange0and2localoptimum}
%\end{figure}

\vspace{0.5cm}
%\begin{figure}[H]
	\centering
  	\includegraphics[width=0.8\textwidth]{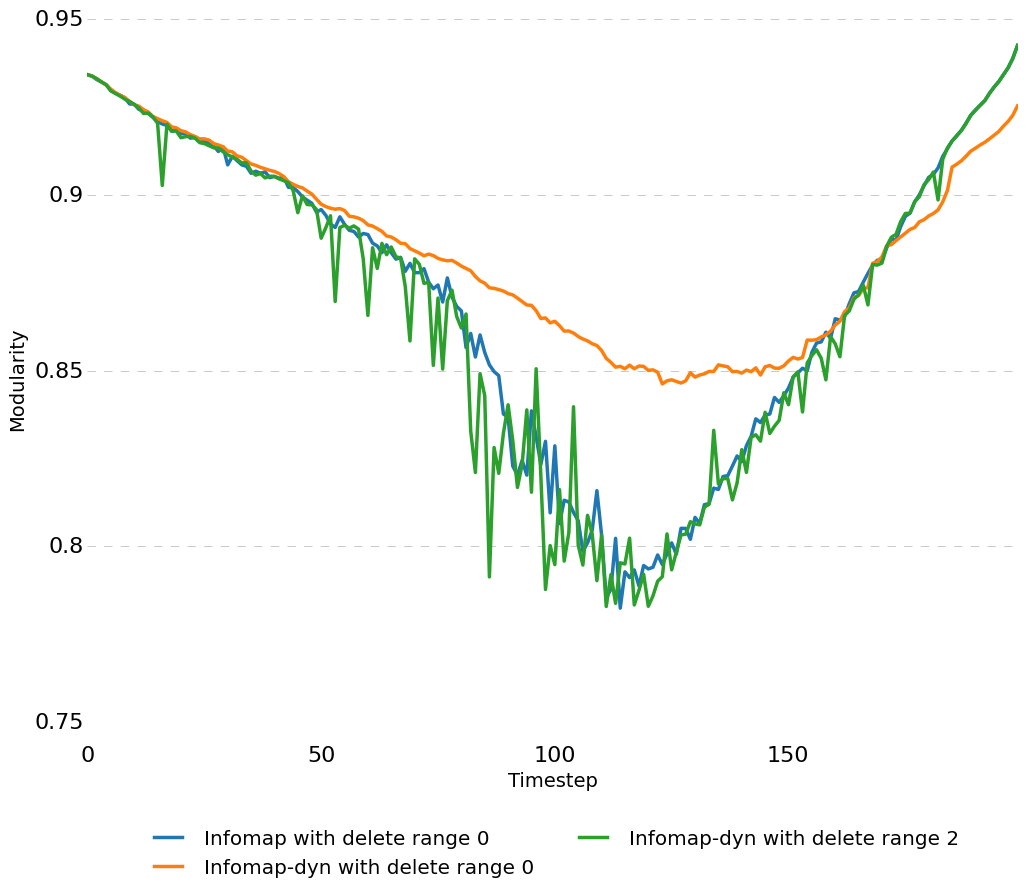}
	\caption{Modularity for Infomap and Infomap-dyn with delete range 0 and 2. Modularity of Infomap-dyn for delete range 0 is even over the values of the static method.}
	\label{fig:deleterange0and2localoptimum-infomapmodularity}
\end{figure}

It was observed that a very low range could cause the algorithm to be trapped in a local optima.
This is due to the fact that the algorithm concentrates on local changes and might not react to more global changes if the \emph{observed} area around changes is too low.
Experiments have shown that a range of at least 2 proves feasible in the chosen scenarios, as seen in \autoref{fig:deleterange0and2localoptimum}.
This can also be seen in \autoref{tab:evaluationvalues}.
The value of \unit[96]{\%} for the Infomap wlogv value of Infomap for delete range 0 can be explained this way.
The dynamic algorithm was not able to escape the local optimum, while this is not the case for higher delete ranges.

The spiky curve that comes with delete range 2 could be explained by the algorithm alternating between two solutions, one being close to the one given by the static method, and one with a lower rating.
With a higher delete range comes a higher time consumption, as already described in Subsection~\ref{sssec:modification}.

\begin{figure}[htb]
	\centering
  	\includegraphics[width=0.8\textwidth]{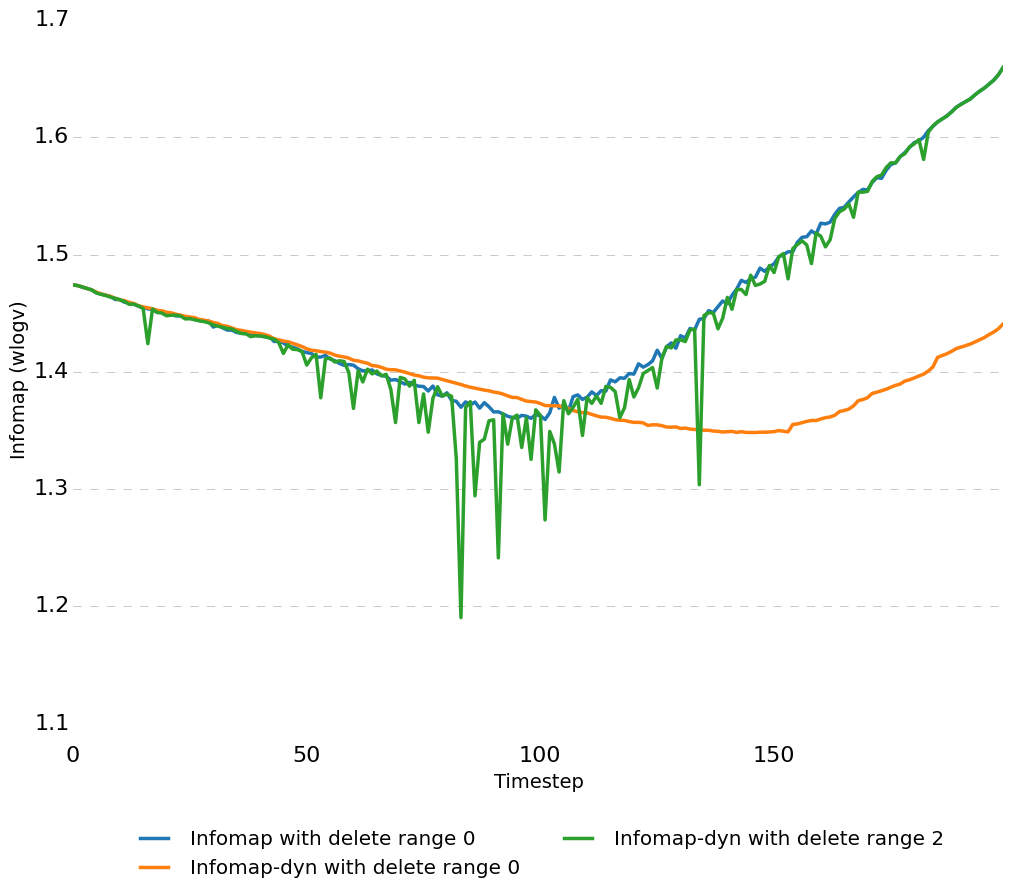}
	\caption{Infomap value (wlogv) for Infomap and Infomap-dyn with delete range 0 and 2. As with Louvain, a local optimum can't be left with delete range 0, but it works with delete range 2.}
	\label{fig:deleterange0and2localoptimum-infomap}
\end{figure}

For the given data, Louvain took \unit[0.72]{s} in average, while Louvain-dyn only took \unit[0.47]{s}.
That is a massive boost in computation time while the local optimum can be overcome and the quality of the clustering stays comparable.
For a delete range of 3 on the other hand, Louvain-dyn took \unit[0.77]{s}.
Therefore it is necessary to estimate, which value for delete range is appropriate and useful for the given task.
The same behavior can be observed with Infomap and the Infomap value.
This can be seen in \autoref{fig:deleterange0and2localoptimum-infomap}.

A curious observation in \autoref{fig:deleterange0and2localoptimum-infomapmodularity} was, that with delete range 0, the dynamic method can actually outperform the static one.
This might be due to the fact that the local optimization is more robust to changes and might adjust better to changing networks.

\section{Conclusion}
\label{sec:conclusion}

In this paper, we present an adaption of the Louvain community detection method to handle changes in dynamic graphs.
We propose to run the standard algorithm only on parts of the graph.
After initialization of the partition, only nodes, which are close to changes, will be processed.
Depending on the definition of \emph{close}, this yields a massive runtime reduction.
Additionally we implemented the Infomap score as an alternative objective function. 

As we are working on dynamic graphs, we provide a graph generator that generates graph steps with strong community structure and intermediate steps that transform one given graph into another one.

In experiments we could show that the dynamic adaption of the algorithm produces similar results with respect to the modularity measure and the Infomap score, with lower calculation time.
In larger graphs local changes affect a much lower percentage of the whole graph, which yields further runtime advantage.

A modularity optimization with a delete range of zero could yield getting stuck in local optima. Higher delete ranges could fix this behavior.  

Future work could transfer the approach presented to an Infomap clustering without the wlogv approximation.
For that task, a formula to efficiently compute the gain of moving a node must be developed. 
Also the influence of the delete range should be investigated for other graph parameters, like very huge graphs with fewer changes per step.

\appendix
\section{Appendix}
\begin{algorithm}[H]
	\algsetup{linenosize=\scriptsize}
  	\scriptsize
	\caption{generate\_dyn\_graph}
	\begin{algorithmic}[1]
		\STATE \textbf{Input:} m, graph\_steps, time\_step\_distance
		\STATE
		\STATE graph\_list $\gets \lbrace \rbrace$
		\STATE last\_graph $\gets$ generate\_graph(m, graph\_steps.pop(0))
		\STATE graph\_list.add(last\_graph)
		\WHILE{\textbar graph\_steps\textbar~\textgreater~0}
			\STATE next\_graph $\gets$ generate\_graph(m, graph\_steps.pop(0))
			\STATE edges\_to\_add, edges\_to\_remove $\gets$ differences(last\_graph, next\_graph)
			\STATE number\_of\_changes $\gets$ \textbar edges\_to\_add\textbar~+~\textbar edges\_to\_remove\textbar
			\STATE changes\_per\_time\_step $\gets \lbrace \rbrace$
			\STATE changes\_per\_time\_step $\gets$ distribute\_evenly(number\_of\_changes, time\_step\_distance)
			\FOR{time\_step := 1 \TO \textbar time\_step\_distance\textbar }
				\STATE current\_graph = graph\_list.get\_last()
				\FOR{change := 0 \TO \textbar changes\_per\_time\_step\textbar }
					\STATE choice $\gets$ random\_choice(add, remove)
					\IF{choice == add}
						\STATE add\_random\_edge\_to\_graph(current\_graph, edges\_to\_add)
					\ELSE
						\STATE remove\_random\_edge\_from\_graph(current\_graph, edges\_to\_remove)
					\ENDIF
				\ENDFOR
				\STATE graph\_list.add(current\_graph)
			\ENDFOR
			\STATE last\_graph $\gets$ next\_graph
		\ENDWHILE
		\RETURN graph\_list
	\end{algorithmic}
	\label{alg:gen_dyn_graph}
\end{algorithm}

\begin{algorithm}[H]
	\algsetup{linenosize=\scriptsize}
  	\scriptsize
	\caption{generate\_graph}
	\begin{algorithmic}[1]
		\STATE \textbf{Input:} m, graph\_step = list of cluster sizes
		\STATE
		\STATE cluster\_list $\gets \lbrace \rbrace$
		\FORALL{cluster\_size \textbf{in} graph\_step}
			\STATE cluster\_list.add(barabasi\_graph(cluster\_size, m))
		\ENDFOR
		\WHILE{\textbar cluster\_list\textbar~\textgreater~ 1}
			\STATE merge\_graph1 $\gets$ cluster\_list.pop(0)
			\STATE merge\_graph2 $\gets$ cluster\_list.pop(0)
			\STATE current\_graph $\gets$ minimal\_merge(merge\_graph1, merge\_graph2)
			\STATE cluster\_list.add(current\_graph)
		\ENDWHILE
		\RETURN current\_graph
	\end{algorithmic}
	\label{alg:gen_graph}
\end{algorithm}

\bibliographystyle{splncs03}
\bibliography{paper}

\end{document}